\documentclass[aps,prb,preprint,floatfix]{revtex4}

\usepackage{amsmath}
\usepackage{graphicx}
\usepackage{enumerate}
\usepackage{dcolumn}

\newcommand{\ddt}[1]{{\frac{\partial {#1}}{\partial t}}}
\newcommand{\ddx}[1]{{\frac{\partial {#1}}{\partial x}}}
\newcommand{\ddvx}[1]{{\frac{\partial {#1}}{\partial v_{x}}}}

\newcommand{\ttt}{{{\bf t}}}
\newcommand{\uu}{{{\bf u}}}
\newcommand{\vv}{{{\bf v}}}

\newcommand{\inv}[1]{{\frac{1}{#1}}}

\newcommand{\ie}{{{\it i.e.}}}
\newcommand{\etal}{{{\it et. al.}}}

\begin{document}

\preprint{APS/123-QED}

\title{Kinetically modified parametric instabilities of 
circularly-polarized Alfv\'en waves:\\
Ion kinetic effects}

\author{Y. Nariyuki}
 \altaffiliation[Also at ]{Department of Earth System Science 
 and Technology, Kyushu University}
\author{T. Hada}%


\date{\today}

\begin{abstract}
Parametric instabilities of parallel propagating, circularly polarized 
finite amplitude Alfv\'en waves in a uniform background plasma is studied, within a 
framework of one-dimensional Vlasov description for ions and massless 
electron fluid, so that kinetic perturbations in the longitudinal direction 
(ion Landau damping) are included. The present formulation also includes 
the Hall effect. 
The obtained results agree well with relevant analysis in the past, 
suggesting that kinetic effects in the longitudinal direction play 
essential roles in the parametric instabilities of Alfv\'en waves when 
the kinetic effects react \lq\lq passively". 
Furthermore, existence of the kinetic parametric instabilities is confirmed 
for the regime with small wave number daughter waves. 
Growth rates of these instabilities are sensitive to ion temperature. 
The formulation and results demonstrated here can be applied to Alfv\'en 
waves observed in the solar wind and in the earth's foreshock region. (accepted to Physics of Plasmas)
\end{abstract}

\pacs{Valid PACS appear here}
\maketitle

Circularly polarized, finite amplitude Alfv\'en waves are ubiquitous in 
collisionless space plasmas, for example in the solar wind, and also in 
the foreshock region of planetary bowshocks where ions backstreaming from 
the bowshock\cite{eastwood05} generate large amplitude Alfv\'en 
waves\cite{agim95,wang03}. 
Parametric instabilities are of particular interest for dissipation of 
quasi-parallel Alfv\'en waves there, since they are typically robust for 
linear ion-cyclotron damping (due to small wave frequencies) and for 
linear Landau damping (due to small propagation angle relative to the 
background magnetic field).

Within a framework of the Hall-magnetohydrodynamic (MHD) equations, 
a number of research has been carried out on the parametric instabilities of 
parallel propagating, circularly polarized finite amplitude Alfv\'en waves\cite{hollweg94,wong86,longtin86,terasawa86}, 
which is an exact solution of the nonlinear Hall-MHD equation sets. 
Among various parametric instabilities, 
modulational, decay, and the beat instabilities are important.
In general, 
growth rates of the instabilities depend
on the parent wave amplitude, wave number, polarization, 
and the ratio of the sound to the Alfv\'en velocity 
($=C_{s}/C_{A}=\sqrt{\beta_{f}}$). 

While a large number of past studies deal with the instabilities 
within the framework of fluid theory, since the ion beta 
(the ion fluid to magnetic pressure ratio) is not small in the 
solar wind and in the earth's foreshock region, the ion kinetic effects
cannot be neglected.
From this perspective, 
numerical analysis using the hybrid simulation code
(super-particle ions + massless electron fluid)
was performed by Terasawa \etal\cite{terasawa86} and 
Vasquez\cite{vasquez95}. 
Mjolhus and Wyller\cite{mjolhus86,mjolhus88} 
derived and discussed in detail a nonlinear evolution 
equation (kinetically modified derivative nonlinear Schrodinger 
equation, first derived by Rogister\cite{rogister71}), 
which describes the wave modulation of weakly despersive, 
nonlinear Alfv\'en waves including the ion Landau damping. 
Fla \etal\cite{fla89} and Spangler\cite{spangler89,spangler90} 
discussed the modulational instability of parallel propagating, 
circularly polarized Alfv\'en waves, and pointed out that,  
while the kinetic effects suppressed the parametric instabilities 
present in the fluid model 
(\lq\lq conservative modulational instability (CMI)"),
the ion kinetics could evoke a new instability 
(\lq\lq resonant particle modulational instability (RPMI)")\cite{fla89}.
Inhester\cite{inhester90} first showed the kinetic formulation for 
parametric decay instabilities without dispersion. 
Gomberoff\cite{gomberoff00} and Araneda\cite{araneda98} discussed the 
ion kinetic effects on parametric instabilities phenomenologically. 
Common view among past analytical works on parametric instabilities 
is that the ion kinetic effects expand the unstable parameter regime 
and reduce the maximum growth rate. 
Recently, Passot and Sulem\cite{passot04} derived the dispersive 
Landau fluid model, and made comparison with some of the other 
models in the past from both analytical and numerical 
standpoints\cite{passot04,bugnon04}. 

Linear analysis play important role in verifying the instabilities seen in the numerical simulation\cite{vasquez95} and provide a more complete reporting of growth rates as a function of the parameters. To our knowledge, linear kinetic analysis with dispersion has 
not yet been demonstrated. 
Our aim in this study is to derive a kinetic formulation including
dispersion systematically, and present a linear analysis and compare it with
other analytical results. 
Our treatment of the ion kinetic effects is similar to that 
in Munoz \etal\cite{munoz05}, in which relativistic 
dispersion relation for parametric instabilities in 
an electron-positron plasma is derived. 

We consider the parametric instabilities of 
parallel propagating, circularly polarized finite amplitude Alfv\'en waves, 
which are the exact solution within the Hall-MHD equation sets.
Assuming weak ion cyclotron damping, we include the kinetic 
effects only along the longitudinal $(x)$ direction.
Let $f(x, t, \vv)$ be the ion distribution function,
and define the integrated longitudinal distribution function, 
\begin{equation}
g(x,t,v_x) = \int f(x,t, \vv) dv_y dv_z .
\label{eq01}
\end{equation}
Then the governing set of equations can be written as, 
\begin{eqnarray}
\ddt{\tilde u} &=& -u_x \ddx{\tilde u} + \inv{\rho} \ddx{\tilde b}, \label{eq02} \\
\ddt{\tilde b} &=& 
- \ddx{} \left( u_x {\tilde b} - {\tilde u} + \frac{i}{\rho} \ddx{\tilde b} \right), 
\label{eq03}\\
\ddt{g} &=& - v_{x}\ddx{g} - (e_{x}+u_{y}b_{z}-u_{z}b_{y})\ddvx{g}, \label{eq04}\\
e_{x} &=& -u_{y}b_{z}+u_{z}b_{y}-\frac{1}{\rho}\ddx{} 
\left( \frac{|b|^{2}}{2}+p_{e} \right), \label{eq05}
\end{eqnarray}
where 
$\rho$ is the plasma density (quasi-neutrality assumed),
$\uu = (u_x, u_y, u_z)$ is the ion bulk velocity vector, 
${\tilde b}=b_y + ib_z$ and ${\tilde u}=u_y + iu_z$ are 
the complex transverse 
magnetic field and bulk velocity, respectively, 
and $e_x$ is the longitudinal electric field. 
All the normalizations have been made using the background constant 
magnetic field, density, Alfv\'en velocity, and the ion gyro-frequency 
defined at a certain reference point. 

The total pressure is given as $p=p_{e}+p_{i}$, 
where isothermal electrons are assumed, \ie, $ p_{e} = T_{e} \rho$.
Also it is assumed that the ion and electron pressures 
are isotropic both at the zeroth and at the perturbation orders. 
Usual beta ratio slightly differs from
the ratio between the sound and the Alfv\'en wave speed squared,
$\beta_{f}=(\gamma_{e}\beta_{e}+\gamma_{i}\beta_{i})/2$, 
where $\gamma_e = 1$ and $\gamma_i$ are the ratios of specific 
heats for electrons and ions, respectively. 

At the zeroth order we consider the  
parallel propagating Alfv\'en wave given as
\begin{eqnarray}
{\tilde b}_p &=& b_0 \exp(i(\omega_0 t -k_0 x)), \label{eq07} \\
{\tilde u}_p &=& u_0 \exp(i(\omega_0 t -k_0 x)), \label{eq08}
\end{eqnarray}
with $u_0 = -b_{0}/v_{\phi 0}$, 
phase velocity $v_{\phi 0}=\omega_{0}/k_{0}$,
$\rho_{0}=1$, $u_{x0}=0$, 
together with the zeroth order dispersion relation
\begin{equation}
\omega_0^2 = k_0^2 (1+\omega_0). \label{eq09}
\end{equation}
We adopt the notation that the positive (negative) 
$\omega_0$ corresponds to the right- (left-) hand polarized waves.
For the zeroth order longitudinal distribution function, we assume
$g_{0}(v_x) =\exp (-v^{2}_{x}/v^{2}_{th})/{\sqrt{\pi}v_{th}}$ 
($v^{2}_{th}=\beta_{i}$).

Then we add small fluctuations given by
\begin{eqnarray}
\delta E &=& E_{+}\exp(i\phi_{+}) + E_{-}\exp(i\phi_{-}), \label{eq10}\\
\delta F &=& \frac{1}{2}(F_{1}\exp(i\Phi) + c.c.), \label{eq11}
\end{eqnarray}
where $E$ represents the transverse variables (${\tilde b}$ and ${\tilde u}$), 
and $F$ represents the longitudinal variables ($g$, $\rho$, and $u_x$), 
$\phi_{\pm}=\omega_{\pm}t-k_{\pm}x$, $\Phi=\Omega t-K x$, 
$k_{\pm}=k_{0} \pm K$, $\omega_{\pm}=\omega_{0} \pm \Omega$, 
and $c.c.$ denotes the complex conjugate.

Then, (\ref{eq02})-(\ref{eq05}) at the first order produce
\begin{eqnarray}
(\omega_{\pm}^{2}-k^{2}_{\pm}-\omega_{\pm}k^{2}_{\pm}) b_{\pm}^{(*)} 
&=& \frac{k_{\pm}}{2}( b_{0}\omega_{\pm}-u_{0}k_{0})u_{x1} \nonumber \\
& &- \frac{b_{0}k_{0}k_{\pm}}{2}( \omega_{\pm} + 1 )\rho_{1}, \label{eq12}
\end{eqnarray}
\begin{eqnarray}
g_{1}(v_x) = -\frac{K[b_{0}(b_{+}+b^{*}_{-})+T_{e}\rho_{1}]}
{\Omega-v_{x}K}\ddvx{g_{0}(v_x)}, \label{eq13}
\end{eqnarray}
where $T_{e}=\beta_{e}/2$ and 
the asterisk denotes the complex conjugate.

Integration of the above yields
\begin{eqnarray}
\rho_{1} &=& -K[b_{0}(b_{+}+b^{*}_{-})+T_{e}\rho_{1}]D, \nonumber \\
D &=& \int_{-\infty}^{\infty}\ddvx{g_{0}}\frac{dv_x}{\Omega-v_{x}K} 
=\frac{2}{v^{2}_{th}K}(1+\xi Z(\xi)), \label{eq14}
\end{eqnarray}
where $\xi =\Omega/v_{th}/K$, 
and $Z(\xi)$ is the plasma dispersion function. 
This equation can be written as
\begin{eqnarray}
\rho_{1} = A(b_{+}+b^{*}_{-}), ~~  A = -\frac{Kb_{0}D}{1+K T_{e}D}.
\label{eq15}
\end{eqnarray}
In a similar way, we have
\begin{eqnarray}
u_{x1} = B(b_{+}+b^{*}_{-}), ~~ B = -K(b_{0}+T_{e}A)C, \label{eq16}\\
C = \int_{-\infty}^{\infty} \ddvx{g_{0}}\frac{v_{x} dv_x}{\Omega-v_{x}K}
=v_{th}\xi D,
\label{eq17}
\end{eqnarray}
where $n_{0}=1$.
Combing (\ref{eq12})-(\ref{eq17}), we obtain
\begin{eqnarray}
L_{+}L_{-}=L_{+}P_{-}+L_{-}P_{+} \label{eq18}
\end{eqnarray}
where
\begin{eqnarray}
L_{\pm} &=& \omega^{2}_{\pm}-k_{\pm}^{2}(1+\omega_{\pm}), \label{eq19}\\
S_{\pm} &=& \frac{b_{0}k_{\pm}}{2}B-\frac{b_{0}k_{0}k_{\pm}}{2}A, \label{eq20}\\
T_{\pm} &=& \frac{v_{0}k_{0}k_{\pm}}{2}B+\frac{b_{0}k_{0}k_{\pm}}{2}A, \label{eq21}\\
P_{\pm} &=& S_{\pm}L_{\pm}-T_{\pm}. \label{eq22}
\end{eqnarray}
By taking the cold limit ($\beta_{i}\sim0$, \ie, $\xi>>1$), 
the dispersion relation of the Alfv\'en wave parametric instabilities 
in the Hall-MHD is obtained\cite{terasawa86}. 
We note that (\ref{eq18}) is an odd function with respect to both $\Omega$ and $K$.


Now we examine the numerical solutions of (\ref{eq18}). 
First, we refer to the results by Inhester\cite{inhester90},
in which the kinetic dispersion relation of parametric decay instability
is derived without dispersion, using drift kinetic model. 
In order to make comparison with their results, we omit the 
dispersion terms in (\ref{eq18}): 
the third term in the RHS of (\ref{eq19}) and the second term
in the RHS of (\ref{eq20}). 
Note that in the dispersionless system, the modulational instability (both CMI and RPMI) does not occur and the growth rate of the decay instability differ from one in the dispersive system\cite{terasawa86}.
By further assuming the cold limit, our equation leads to 
the MHD dispersion relation obtained by 
Goldstein\cite{goldstein78} and Derby\cite{derby78},
(but not to the model by Inhester (1990)).
On the other hand, numerically obtained dispersion relations 
based on our model and that of Inhester at least qualitatively 
agrees, in the sense that the ion kinetic effects enlarges the 
unstable parameter regime and also 
the maximum growth rates are reduced (Figure 1). 
These results are also in qualitative agreement with 
some of the past works\cite{inhester90,bugnon04}.

%
%
\begin{figure}
\noindent\includegraphics[width=18pc]{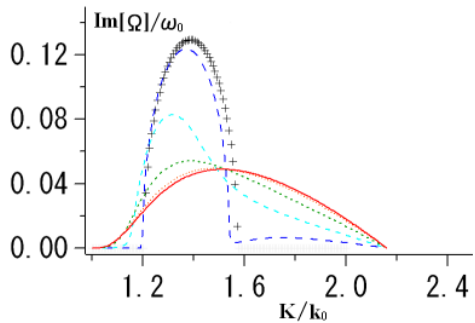}
\caption{
Growth rates of the density modes 
(whose frequency/wave number is normalized to the parent wave 
parameters), driven by the decay instability of dispersionless 
Alfv\'en waves with amplitude $b_{0}=0.447$, $\beta=0.6$, 
and $T_{r}(=T_{i}/T_{e})=$ (a) $0$ (=fluid model, black cross), 
(b) $0.03$ (blue broken line), (c) $0.2$ (light blue), 
(d) $1$ (green dotted line), (e) $5$ (orange), 
(f) $\infty (T_{e}=0)$ (red solid line). 
The results shown here are in agreement with 
Inhester (1990) and Bugnon \etal (2004). }
\end{figure}
We now turn our attention to the parametric instabilities of the 
dispersive Alfv\'en waves. 
Figure 2 shows the growth rates computed from (\ref{eq18}) 
for right-hand polarized Alfv\'en waves, for 
various temperature ratios, $T_{r}(=T_{i}/T_{e})$. 
The decay-like instabilities have positive
growth rates when $K>|k_{0}|=0.408$. 
The results here qualitatively agrees with numerical results 
obtained by Vasquez\cite{vasquez95} and Bugnon\cite{bugnon04} 
(Table 1 in both papers), which suggests that 
the ion kinetic effects in the longitudinal direction
play essential roles in kinetic modification of the Alfv\'en 
parametric instabilities, when 
$\beta_{i}$ is relatively small, \ie, the kinetic effects 
\lq\lq passively" 
react to the fluid dynamics. 
%
%
\begin{figure}
\noindent\includegraphics[width=18pc]{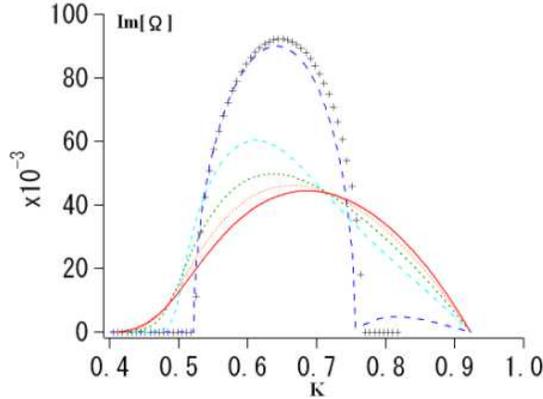}
\caption{Growth rates of the density modes 
(frequency/wave number is normalized to the
ion-gyrofrequency/ion inertial length), 
driven by the decay instability of dispersive 
Alfv\'en waves with 
$b_{0}=0.5$, $k_{0}=0.408$, $\omega_{0}=0.5$ (R-mode), 
$\beta=0.45$, and $T_{r}(=T_{i}/T_{e})=$ 
(a) $0$ (=fluid model, black cross), (b) $0.0227$ (blue broken line), 
(c) $0.36$ (light blue), (d) $1$ (green dotted line), (e) $2.75$ (orange), 
(f) $\infty (T_{e}=0) $(red solid line). 
The results agree well with Vasquez (1995) and Bugnon (2004). }
\end{figure}
Figure 3 shows the growth rates under the same parameters as in Figure 2
except that $k_{0}=0.102$ and $\omega_{0}=0.107$ (R-mode). 
Both regimes $K>|k_{0}|$ and $K<|k_{0}|$ are plotted. 
The former corresponds to the \lq\lq conservative decay instability (CDI)",
which has the finite growth rate at $T_{r}=0$ as we see in Figures 1 and 2. 
On the other hand, the latter, the RPMI,
is destabilized only for finite $T_{r}$, although the growth rate is 
typically 1 or 2 orders less than the CMI. 
The RPMI is quenched when $T_{r}=0$, suggesting that 
the instability is a product of Landau resonant effects
(Fla \etal\cite{fla89}, Spangler\cite{spangler89,spangler90}). 
The RPMI exists even at small wavenumbers.
%
%
\begin{figure}
\noindent\includegraphics[width=18pc]{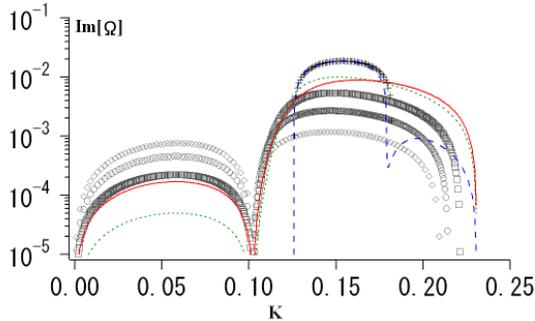}
\caption{
Same as Figure 2 except that
$b_{0}=0.5$, $k_{0}=0.102$, $\omega_{0}=0.107$ (R-mode), $\beta=0.45$, 
with $T_{r}(=T_{i}/T_{e})=$ (a) $0$ (=fluid model, black cross), 
(b) $0.0227$ (blue broken line), (c) $1 (\beta_{e}=\beta_{i}=0.225)$ 
(green dotted line), (d) $\infty (T_{e}=0)$ (red solid line), 
and when $\beta_{e}=0.225$ with (e) $\beta_{i}=0.5$ (square), 
(f) $\beta_{i}=1.0$ (circle), (g)$\beta_{i}=2.0$ (diamond), 
respectively.}
\end{figure}

As shown in Figure 3 (c)-(g), the growth rate of the RPMI is sensitive to 
$\beta_i$. 
In particular, as $\beta_i$ is increased, the growth rate of the RPMI is
enhanced while that of the CDI is reduced. 
When $\beta_{i}=2$ and $\beta_{e}=0.225$, the growth rates of 
both instabilities become comparable. 
The CDI growth rate is increased as $\beta_f$ (=$\beta_e/2$ for the present case) 
is reduced. 
The decrease of the growth rates of the CD(M)I is 
relaxed at large $T_{r}$ ({\it c.f.}, Fig.1, 2 (e), (f)).

Finally, we discuss briefly the instability of left-hand polarized Alfv\'en waves
(Figure 4). 
Under the parameters used, the CMI (the maximum growth rate $\sim 0.4$), 
the CDI (likewise, $\sim 0.6$), and the beat instability (likewise, $\sim 0.8$) are driven 
unstable at $T_{r}=0$. 
Figure 4 shows that the growth rate of the decay-like instability ($\sim 0.8$)
exceeds that of the modulational-like instability at $T_{r}> \sim 0.2$.

%
%
\begin{figure}
\noindent\includegraphics[width=18pc]{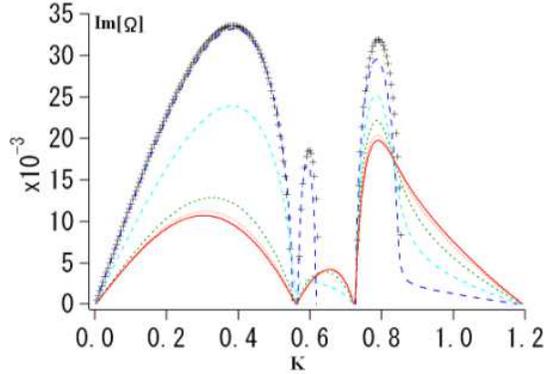}
\caption{Same as Figure 2 except that
$b_{0}=0.4$, $k_{0}=0.56$, $\omega_{0}=-0.425$ (L-mode), $\beta=0.6$, 
and $T_{r}(=T_{i}/T_{e})=$ (a) $0$ (=fluid model, black cross), 
(b) $0.03$ (blue broken line), (c) $0.2$ (light blue), 
(d) $1$ (green dotted line), (e) $5$ (orange), and
(f) $\infty(T_{e}=0)$ (red solid line). }
\end{figure}

In the present paper we 
discussed the kinetically modified parametric instabilities
of circularly polarized, parallel-propagating Alfv\'en waves 
within a framework of one-dimensional system, which includes 
longitudinal kinetic perturbations. 
The obtained dispersion relation (\ref{eq18}) is numerically evaluated, 
and compared with relevant works in the literature.
The real part of $\Omega$ in (\ref{eq18}) is not shown here, but
we have confirmed that the propagation direction of the sideband 
Alfv\'en waves and the density fluctuations are consistent with 
the past studies. 

While our analysis includes $\beta_i$ as a free parameter, 
it should be retained at a modest value,
since at high $\beta_i$, kinetic response of the transverse
distribution function cannot be neglected, and the kinetic
effects must be included in the parent wave dispersion relation
(9)\cite{vasquez95, bugnon04, abraham77, yajima66}. 
Comparison of Fig.2 with past studies suggests that our 
assumption on basic equations is valid when $\beta_{i}=0.45$,
but is not for $\beta_{i} \sim>1.5$\cite{vasquez95}. 
This remark also applies to simulation studies using the
hybrid code: it is important to use kinetic instead of fluid
dispersion relation to give initial wave when the ion beta is large.

Finally we point out that, despite the importance of the parametric
instabilities in the solar wind or in the earth's foreshock, 
their presence has not yet been clearly demonstrated by spacecraft 
experiments.
Linear theory predicts that parallel propagating waves are mainly 
excited in the solar wind or in the earth's foreshock\cite{gary81, hada87}. 
However, recent multi-point measurement observations suggest that 
these waves tend to propagate obliquely to the background magnetic 
field\cite{eastwood05, narita03, eastwood05b}. 
Numerical simulation study using hybrid code\cite{wang03} should be 
effective in understanding the parametric instabilities of obliquely propagating 
large amplitude MHD waves. 
We hope the situation to change via knowledge of kinetic modification 
of the instabilities as well as state-of-the-art data acquisition 
and analysis techniques. 

We thank Drs. S. Matsukiyo, V. Munoz, and T. Passot for fruitful discussions 
and comments. 
This paper has been supported by JSPS Research Fellowships for 
Young Scientists in Japan.


\begin{thebibliography}{}
\bibitem{eastwood05} J. P. Eastwood, , E. A. Lucek, 
C. Mazelle, K. Meziane, Y. Narita, J. Pickett, and R. A. Treumann, 
 Space. Sci. Rev \textbf{118}, 41 (2005).
\bibitem{agim95}  Y. Z. Agim, A. F. Vinaz, and M. L. Goldstein, 
 J. Geophys. Res \textbf{100}, 17081 (1995).
\bibitem{wang03} X. Y. Wang, Y. Lin, Phys. Plasmas \textbf{10}(9), 3528 (2003).
\bibitem{hollweg94} J. V. Hollweg, 
 J. Geophys. Res \textbf{99}, 23431 (1994).
\bibitem{wong86}H. K. Wong, M. L. Goldstein, 
 J. Geophys. Res \textbf{91}, 5617 (1986).
\bibitem{longtin86} M. Longtin, B. U. O. Sonnerup, 
 J. Geophys. Res \textbf{91}, 798 (1986).
\bibitem{terasawa86} T. Terasawa, M. Hoshino, J. -I. Sakai, T. Hada,
 J. Geophys. Res \textbf{91}, 4171 (1986).
\bibitem{vasquez95} B. J. Vasquez, 
 J. Geophys. Res \textbf{100}, 1779 (1995).
\bibitem{mjolhus86} E. Mj{\o}lhus, J. Wyller, 
 Phys. Scr \textbf{33}, 442 (1986).
\bibitem{mjolhus88} E. Mj{\o}lhus, J. Wyller, 
 J. Plasma Phys \textbf{40}, 299 (1988).
\bibitem{rogister71} A. Rogister, 
 Phys. Fluids \textbf{14}, 2733 (1971).
\bibitem{fla89} T. Fla, E. Mj{\o}lhus,  and J. Wyller, 
 Phys. Scr \textbf{40}, 219 (1989).
\bibitem{spangler89} S. R. Spangler, 
 Phys. Fluids \textbf{B1} (8), 1738 (1989).
\bibitem{spangler90} S. R. Spangler,
 Phys. Fluids \textbf{B2} (2), 407 (1990).
\bibitem{inhester90} B. Inhester, J. Geophys. Res \textbf{95}, 10525 (1990).
\bibitem{gomberoff00} L. Gomberoff, J. Geophys. Res \textbf{105}, 10509 (2000).
\bibitem{araneda98} J. A. Araneda, Phys. Scr \textbf{75}, 164 (1998).
\bibitem{passot04} T. Passot, P. L. Sulem, 
 Phys. Plasmas \textbf{11} (11), 5173 (2004).
\bibitem{bugnon04} G. Bugnon, T. Passot, P. L. Sulem,
 Nonl. Proc. in Geophys \textbf{11}, 609 (2004).
\bibitem{munoz05} V. Munoz, T. Hada, S. Matsukiyo, Earth Planets Space \textbf{58}, 1213, 2006.
\bibitem{goldstein78} M. L. Goldstein, 
 Astrophys. J \textbf{219} (2), 700 (1978).
\bibitem{derby78} N. F. Derby, 
 Astrophys. J \textbf{224} (3), 1013 (1978).
\bibitem{abraham77} B. Abraham-Shrauner, W. C. Feldman, 
J. Geophys. Res \textbf{82}, 618 (1977).
\bibitem{yajima66} N. Yajima, Prog. Theor. Phys. \textbf{36}(1), 1 (1966).
\bibitem{gary81} S. P. Gary, J. T. Gosling, D. W. Forslund, J. Geophys. Res \textbf{86}, 6691 (1981).
\bibitem{hada87} T. Hada, C. F. Kennel, T. Terasawa, J. Geophys. Res \textbf{92}, 4423 (1987).
\bibitem{narita03} Y. Narita, K. -H. Glassmeier, S. Schafer, U. Motschmann, S. Sauer, I. Dandouras, K. -H. Fornacon, E. Georgescu, H. Reme, Geophys. Res. Lett \textbf{30}(13), 1710 (2003).
\bibitem{eastwood05b} J. P. Eastwood, A. Balogh, E. A. Lucek, C. Mazelle, I. Dandouras, J. Geophys. Res \textbf{110}, 11220 (2005).
\end{thebibliography}
\end{document}